\documentclass[
prb,
aps,
onecolumn,
twocolumn,
amsmath,
amssymb,
superscriptaddress,
floatfix
]{revtex4-2}
\usepackage[T1]{fontenc}
\usepackage[english]{babel}
\usepackage{graphicx}
\usepackage{dcolumn}
\usepackage{amsmath}
\usepackage{amssymb}
\usepackage{bm}
\usepackage{multirow}
\usepackage[normalem]{ulem}
\usepackage[usenames,dvipsnames,svgnames,table]{xcolor}
\usepackage{braket}
\usepackage{subfigure}
\usepackage[colorlinks]{hyperref}
\usepackage{physics}

\newcommand{\refeqn}[1]{(\ref{#1})} 
\newcommand{\Hamil}{\mathcal{H}} 
\newcommand{\vect}[1]{\textbf #1}

\renewcommand{\exp}[1]{\text{e}^{#1}}
\renewcommand{\i}{\text{i}}

\usepackage{accents}
\makeatletter
\DeclareRobustCommand{\cev}[1]{%
  \mathpalette\do@cev{#1}%
}
\newcommand{\do@cev}[2]{%
  \fix@cev{#1}{+}%
  \reflectbox{$\m@th#1\vec{\reflectbox{$\fix@cev{#1}{-}\m@th#1#2\fix@cev{#1}{+}$}}$}%
  \fix@cev{#1}{-}%
}
\newcommand{\fix@cev}[2]{%
  \ifx#1\displaystyle
    \mkern#23mu
  \else
    \ifx#1\textstyle
      \mkern#23mu
    \else
      \ifx#1\scriptstyle
        \mkern#22mu
      \else
        \mkern#22mu
      \fi
    \fi
  \fi
}
\makeatother

\begin{document}

\title{
Josephson tunneling through a Yu-Shiba-Rusinov state: Interplay of $\pi$-shifts in Josephson current and local superconducting order parameter  
}
\author{Andreas Theiler}
 \affiliation{Department of Physics and Astronomy, Uppsala University, Box 516, 
751 20 Uppsala, Sweden}

\author{Christian R. Ast}
\affiliation{Max-Planck-Institut f\"ur Festk\"orperforschung, Heisenbergstra\ss e 1, 70569 Stuttgart, Germany}

\author{Annica M. Black-Schaffer}
\affiliation{Department of Physics and Astronomy, Uppsala University, Box 516, 
751 20 Uppsala, Sweden}

\date{\today}

\begin{abstract}
An impurity hosting a magnetic moment coupled to a conventional $s$-wave superconductor gives rise to so-called Yu-Shiba-Rusinov (YSR) states with energies inside the superconducting gap.
Depending on the coupling between the impurity and the superconductor, the system can have two distinct quantum ground states separated by a quantum phase transition (QPT).
We investigate the interplay of two effects observed at the QPT. 
First, the tunneling supercurrent through the impurity reverses its sign at the QPT, denoted as a $\pi$-shift in the current-phase relation.
Secondly, the local superconducting order parameter at the impurity site is suppressed and becomes negative at the QPT, generally termed a $\pi$-shift in the local superconducting order parameter. 
We find that both these effects are governed by the presence of the YSR state, however, they do not significantly depend or influence each other.
In particular, we establish that the $\pi$-shift in the superconducting order parameter does not induce a $\pi$-shift in the tunneling Josephson current, nor can the Josephson current and its spatial behavior be used to directly probe the impurity-induced changes in the local superconducting order parameter, which occur on a length scale substantially shorter than the superconducting coherence length.
\end{abstract}

\maketitle

\section{Introduction}
\label{sec:introduction}

Magnetic impurities coupled to superconductors give rise to localized in gap states, so-called Yu-Shiba-Rusinov (YSR) states \cite{luh1965bound, shiba_classical_1968, rusinov1969superconductivity, rusinov_theory_1969}.
These states have numerous interesting properties and have also been used to probe properties of superconducting systems, such as the Fermi-surface \cite{ortuzar_yu-shiba-rusinov_2022, trivini_diluted_2024}, quasi-particle interference  \cite{zaldivar_revealing_2025} or allow for Josephson tunneling experiments at the atomic limit \cite{huang2020tunnelling}.
They have further been proposed as building blocks for qubits \cite{mishra_yu-shiba-rusinov_2021, pavesic_qubit_2022,  steffensen_ysr_2024} and for creating systems hosting Majorana zero mode systems \cite{nadj-perge_observation_2014, ruby_end_2015, kezilebieke_coupled_2018, ruby_wave-function_2018, zatelli_robust_2024}, encouraging further exploration of their properties.

Depending on the coupling between the magnetic impurity and the superconductor, the system can go through a quantum phase transition (QPT).
In the weak coupling regime, the magnetic impurity is only weakly screened by the superconducting condensate, forming a spin-doublet ground state (assuming a spin-$\frac{1}{2}$ impurity), while at strong coupling the YSR state becomes occupied, with an electron fully screening the impurity spin, forming a spin-singlet ground state \cite{balatsky_impurity-induced_2006}.
The transition between these two ground states occurs at a critical coupling that marks the QPT.
In the energy spectrum, the YSR state can be observed as two distinct spectral features within the superconducting energy gap, symmetric around the Fermi level.
At the QPT, these two YSR states cross through zero energy, and thus the occupation between the two states switches. 

The QPT has been observed in scanning tunneling microscopy (STM) and Josephson-STM (JSTM) experiments \cite{ternes2006scanning, farinacci_tuning_2018, brand2018electron, malavolti_tunable_2018, kezilebieke_observation_2019, huang_quantum_2020, karan_superconducting_2022, huang_universal_2023, karan_tracking_2024, uldemolins_hunds_2024}.
A typical setup for such an experiment is to place a magnetic impurity either on a superconducting substrate or a superconducting tip, while manipulating the impurity via interatomic forces by adjusting the tip-sample distance.
This allows for measuring the Josephson tunneling current through the magnetic impurity at various coupling strengths between the impurity and the superconductor.

In a generic Josephson superconductor-superconductor (S|S) tunnel junction, the Josephson current at zero-bias voltage varies with the phase difference $\varphi$ between the two superconductors.
The maximum of the Josephson current is the critical current $I_C$, which in ideal junctions occurs at $\varphi=\pi/2$. 
However, for a S|magnetic impurity|S junction, the superconducting phase difference at which the maximum Josephson current occurs has been shown to differ by $\pi$, i.e.~$I_C$ switches sign, depending on whether the system is in the strong or weak coupling regime \cite{vecino_josephson_2003, dam_supercurrent_2006, martin2011josephson, graham2017imaging, karan_tracking_2024}.
Therefore, there is a transition in the behavior of the Josephson current at the QPT, denoted as a $\pi$-shift in the Josephson current.
This $\pi$-shift can be challenging to detect in experiments, as the critical Josephson current is only measured through the switching current $I_S$, which only gives information about the absolute value $|I_C|$. Still, by utilizing that electrons can  also tunnel through different transport channels into the substrate that do not involve the impurity orbital that hosts the magnetic moment, it has been possible to experimentally demonstrate this current $\pi$-shift \cite{karan_superconducting_2022}.

The presence of the YSR impurity state also dramatically decreases the local superconducting order parameter in the superconductor, as the magnetism of the impurity and conventional superconductivity are antagonistic effects \cite{rusinov1969superconductivity,heinrichs1968spatial, schlottmann1976spatial, kim1993spatial, balatsky_impurity-induced_2006}.
At the QPT, the superconducting order parameter has even been shown to locally become negative at the impurity site in the strong coupling regime \cite{salkola1997spectral, flatte1997localDefects, flatte1997localMagnetic, balatsky_impurity-induced_2006, meng2015superconducting, pershoguba2015currents, graham2017imaging, bjornson2017superconducting, theiler2019majorana, theiler_temperature_2025}. 
This is typically referred to as a $\pi$-shift as well, but in the local superconducting order parameter.

In this work, we explore whether the suppression and $\pi$-shift of the superconducting order parameter influence the Josephson tunneling current, such that it would be detectable in JSTM experiments, and, in particular whether the $\pi$-shift in the Josephson current is related to the $\pi$-shift in the local superconducting order parameter.
The existence of such a relationship has been proposed before~\cite{graham2017imaging}, based on the development of similar spatial patterns of the Josephson current and the local order parameter.
In contrast, we have previously discussed similar results \cite{karan_superconducting_2022}, but then found no direct connections between the two $\pi$-shifts.
Here we substantially expand on these results by providing a more detailed description for the magnetic impurity and the tunneling current, including more tunneling channels, and we also highlight the effects of temperature on the system.

\begin{figure}
    \centering
    \includegraphics[width=0.49\linewidth]{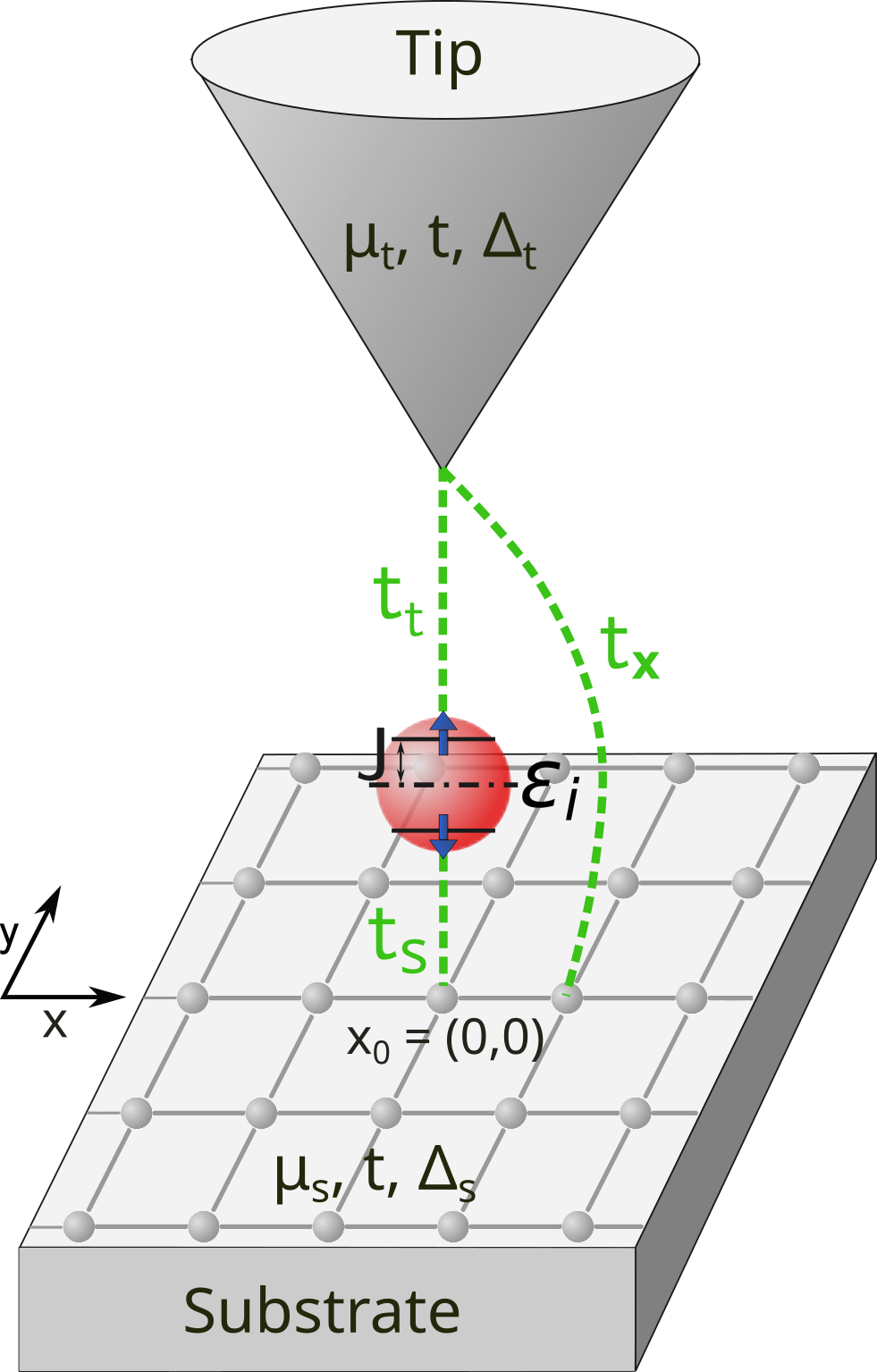}%
    \caption{Schematic setup.
    The magnetic impurity, responsible for creating the YSR states is depicted in red and is defined by the onsite energy level $\epsilon_i$ and spin-dependent energy splitting level interaction $J$.
    The impurity is coupled to the STM tip (top) through a hopping term $t_t$ and to the substrate by $t_s$ to the lattice site $x_0$.
    Additionally, there are other possible transmission channels between the tip and substrate with couplings $t_\vect{x}$.
    }
    \label{fig:toymodel}
\end{figure}

The remainder of this work may be summarized as follows.
In Section \ref{subsec:physical_model} we introduce the model of the system, consisting of a magnetic impurity absorbed on a superconducting substrate and with a superconducting tip, forming a tunnel junction, see Fig.~\ref{fig:toymodel}.
Section \ref{subsec:method_order_parameter} addresses how we calculate the suppression of the superconducting order parameter in the substrate as a function of temperature, while Sections \ref{subsec:josephson_current_through_imp_methods} and \ref{subsec:methods_substrate_tunneling} show how the Josephson current is calculated. 
In Section \ref{subsec:energy_spectrum} we report our results for tunneling through the impurity.
In particular, we show that the Josephson current through the impurity only experiences very minor changes when the suppression of the local order parameter is taken into account. We predict that these changes will be challenging to detect in experiments. As a consequence, the local order parameter $\pi$-shift does not cause the current $\pi$-shift. In Section \ref{subsec:results_scanning_stm} we further report on the spatial dependence of the Josephson tunneling and show again that the suppression of the local order parameter has a negligible influence on the Josephson current.
Furthermore, we show that, in the absence of a YSR state, but still artificially imposing a $\pi$-shift in the local superconducting order parameter, this artificial $\pi$-shift does not induce a $\pi$-shift in the Josephson current. This further establishes that the current $\pi$-shift is due to the YSR state itself and not due to the induced $\pi$-shift in the local order parameter.
Finally, we add additional transmission channels in Section \ref{subsec:results_additonal_channels} and show that even transmission channels that avoid the impurity site can still be modified by the YSR state, and thereby can exhibit a $\pi$-shift in the Josephson current, or at least a discontinuity without a sign change. 

Overall, our results establish that the sign change in the Josephson current through a YSR state at the QPT is not due to the $\pi$-shift experienced by the local superconducting order parameter at the QPT. As a consequence, JSTM is not a reliable tool for probing the local superconducting order parameter around a magnetic impurity.

\section{Modeling and methods}

For modeling the magnetic impurity, we utilize the Anderson impurity model in the mean field description and couple it to a conventional $s$-wave superconductor and a superconducting STM tip, both described by a Bogoliubov-de~Gennes (BdG) tight binding model, see Figure~\ref{fig:toymodel} and Section~\ref{subsec:physical_model} for more details.
We then derive the critical Josephson current for Cooper pairs tunneling from the tip to the impurity in Section \ref{subsec:josephson_current_through_imp_methods}, to the substrate in Section \ref{subsec:methods_substrate_tunneling}, and add more transmission channels for the tunneling current in Section \ref{subsec:extra_channels_methods}.

\subsection{Tip-impurity-substrate system}
\label{subsec:physical_model}

We focus on a general model that describes the magnetic impurity as an Anderson impurity that is coupled to two superconducting leads that represent the superconducting substrate and the superconducting tip separately, see Fig.~\ref{fig:toymodel}.
The Hamiltonian of the combined system is then:
\begin{equation}
    \Hamil = \Hamil_{i} + \sum_{\alpha=s,t} \left( \Hamil_{\alpha} + V_{i,\alpha} \right)
\end{equation}
where $\Hamil_{i}$ describes the impurity and $\Hamil_{\alpha}$ are the Hamiltonians for the superconducting substrate and tip, with indices $s$ and $t$, respectively.
These three system components are connected through couplings $V_{i,\alpha}$.

\subsubsection{Tip and substrate}

The superconducting tip and substrate are described by two BdG tight-binding Hamiltonians in a real space:
\begin{equation}
    \label{eqn:Hamil_subsystems}
    \begin{split}
    \Hamil_{\alpha} &= \mu_{\alpha} \sum_{\vect{i}, \sigma, \sigma} c^{\dagger}_{\vect{i} \sigma, \alpha} c_{\vect{i} \sigma, \alpha} - t_{0,\alpha} \sum_{\langle \vect{i},\vect{j} \rangle , \sigma} c^{\dagger}_{\vect{i} \sigma, \alpha} c_{\vect{j} \sigma, \alpha} \\
    &+ \sum_{i} \Delta_{\vect{i}, \alpha} \exp{\i \varphi_\alpha} c_{\vect{i} \uparrow, \alpha} c_{\vect{i} \downarrow, \alpha} + \text{h.c.},  
    \end{split}
\end{equation}
where the indices $\langle \vect{i},\vect{j} \rangle$ describe the sum over all nearest neighbors on a two dimensional square lattice.
The local superconducting order parameter is defined by $\Delta_{\vect{i}, \alpha}$ for each lattice site $\vect{i}$, with its phase given by $\varphi_\alpha$ being a constant throughout each subsystem.
For a homogeneous system, or for sites far away from perturbations such as an impurity, the site-dependent order parameter is constant, and we denote its value by the bulk order parameter $\Delta_{0, \alpha}$.
The remaining parameters $\mu_\alpha$ and $t_{0,\alpha}$ denote the chemical potential and the nearest neighbor hopping, respectively.
The hopping energy $t_{0,\alpha}$ defines the bandwidth of the normal state band structure, which is $8\,t_{0,\alpha}$ for the square lattice.

We simulate the tip Hamiltonian $\Hamil_t$ and substrate Hamiltonian $\Hamil_s$ using the same model parameters, i.e. $t_{0,s} = t_{0,t} = t$, $\mu_s = \mu_t =\mu$ and $\Delta_{0,s} = \Delta_{0,t} = \Delta_{0}$.
This assumption is based on common experimental setups, which generally use the same material for the tip and substrate, with the same or similar superconducting gaps in both leads \cite{farinacci_tuning_2018, kezilebieke_coupled_2018, huang_quantum_2020, karan_superconducting_2022, huang_universal_2023, karan_tracking_2024}.
This indicates that both have a similar density of states at the Fermi level, set by $t$ and $\mu$ and a similar superconducting interaction strength, resulting in the same bulk order parameter $\Delta_{0}$.
We set the hopping parameter $t$ to unity and use it as the energy reference for all band structure related parameters.
The chemical potential is set to $\mu_\alpha = 0.5\,t$ to avoid the van Hove singularity, as we are interested in using a generic band structure.

Both the tip and the substrate are calculated using periodic boundary conditions to avoid artificial boundary effects and use a grid size of $111 \times 111$, which avoids finite size related effects due to the impurity for all results that we present in this work.
Furthermore, we tested a one dimensional geometry for the tip and different choices for the chemical potential $\mu$, and found that these choices did not qualitatively change our results.

\subsubsection{Magnetic impurity}

We model the magnetic impurity using the Anderson impurity model within the mean-field description, where the onsite Coulomb repulsion results in an effective energy separation $J$ between the two spin orientations, and where we treat $J$ as a free parameter.
This mean-field description has been shown to capture key aspects of the quantum impurity problem \cite{martin2011josephson, martin-rodero_andreev_2012, zitko_shiba_2015, kadlecova2017QuantumDot, zitko_quantum_2018, kadlecova2019PracticalGuide}. Further, a fully interacting quantum impurity model requires computationally demanding techniques, such as numerical renormalization group \cite{wilson1975NRG} or quantum Monte Carlo methods \cite{RevModPhys.83.349, PhysRevB.81.024509, PhysRevLett.108.227001}, which become too numerically costly as we also need to self-consistently compute the superconducting order parameter in the substrate, see \ref{subsec:method_order_parameter}.
The mean-field Anderson impurity Hamiltonian for the impurity reads \cite{huang_quantum_2020}:
\begin{equation}
\label{eqn:hamil_mag_imp}
         \Hamil_{i} =  \sum_{\sigma, \sigma'} \left( \epsilon_i \sigma_0 + \vect{J} \vec{\mathbf{\sigma}} \right)_{\sigma, \sigma'} d^{\dagger}_{\sigma} d_{\sigma'},
\end{equation}
where $d^{(\dagger)}_\sigma$ are the fermionic annihilation (creation) operators for the impurity with spin $\sigma$.
The onsite energy level of the impurity is $\epsilon_i$, which includes the renormalization of the non-interaction energy levels caused by the mean-field treatment of the Coulomb repulsion.
The effective magnetic energy level splitting for opposite spin orientations is denoted by $\vect{J}$, which is generated by the Coulomb interaction at the impurity site producing a local magnetic moment on the impurity.
Without loss of generality, we choose $\vect{J} = J \hat{z}$. We further set $J = 1\,t$ and keep it constant in our calculations, as we use the coupling between the impurity and the superconductors as the tuning parameter. 
Although the interaction between impurity and other superconductors can in principle renormalize $J$ via self-energy effects, we assume that these effects are minor.
In addition, we use $\epsilon_i = 0$ throughout this work.
While it is realistic to expect an impurity to have a finite onsite potential, we find that the choice of $\epsilon_i$ does not change the results qualitatively.

\subsubsection{Couplings between tip, impurity, and substrate}

The coupling between the impurity and the substrate or tip is described by:
\begin{equation}
\label{eqn:Hamil_coupling_general}
    V_{i,\alpha} = - \sum_{\sigma} t_{\alpha}  d^{\dagger}_{\sigma} c_{\vect{x}_0, \sigma,  \alpha} + \text{h.c.},
\end{equation}
with $t_{\alpha}$ denoting the hopping between the impurity and the superconductor $\alpha$ and $\vect{x}_0$ is the site index in the subsystem $\alpha$ to which the impurity is coupled.
To model a typical JSTM experiment setup, we assume that there is a weak tunneling coupling between the impurity and the STM tip, while the coupling between the impurity and the substrate is much stronger, i.e. $t_t \ll t_s$.
Some experiments instead functionalize the superconducting tip with a magnetic impurity and use a clean substrate \cite{huang_quantum_2020, karan_superconducting_2022, karan_superconducting_2022, huang_universal_2023, karan_tracking_2024}, in this case the roles of tip and substrate are simply reversed.
As the coupling between the impurity and the tip describes a tunnel junction, we set it a low value at $t_t = 0.05\,t$ and keep it constant for simplicity.
The exact value of $t_t$ does not change our results qualitatively, since it is only a prefactor in the calculation of the critical Josephson current in Eq.~\refeqn{eqn:dc_josephson_current}.

\subsubsection{YSR state energy spectrum}

Ignoring the effects of the local order parameter suppression in the substrate, the coupling to the tip, the energy of the YSR states is given by \cite{villas2020interplay, huang_quantum_2020, karan_superconducting_2022}:
\begin{equation}
    \varepsilon_{YSR} = \pm \Delta_{0} \frac{J^2 - \Gamma_{s}^2 - \epsilon_i^2 }{\sqrt{\left(\Gamma_{s}^2 + \left( J + \epsilon_i \right)^2 \right) \left(\Gamma_{s}^2 + \left( J - \epsilon_i \right)^2 \right) }},
\end{equation}
where $\Gamma_{s} = N_{0,s} \pi t_s^2$ denotes the coupling rate between the impurity and $N_{0,s}$ denotes the substrate normal state density of states at the Fermi level.
At the QPT, the YSR state energies cross the Fermi surface and therefore change occupation \cite{balatsky_impurity-induced_2006}.
This occurs at the critical coupling rate $\Gamma_{C}^2 = J^2 - \epsilon_i^2$, which yields for the critical coupling constant $t_C = \sqrt{\tfrac{\Gamma_C}{ N_{0,s} \pi}}$.
Thus, the system can be driven through the QPT by varying $t_{s}$. Experimentally, $t_s$ can also be changed, as it is determined by the distance between the impurity and the substrate and, therefore, can be controlled via atomic forces when moving the tip \cite{ternes2006scanning, farinacci_tuning_2018, brand2018electron, malavolti_tunable_2018, kezilebieke_observation_2019, huang_quantum_2020, karan_superconducting_2022, huang_universal_2023, karan_tracking_2024}.

%
\subsection{Local superconducting order parameter}
\label{subsec:method_order_parameter}

The presence of a magnetic impurity close to a superconductor suppresses the local superconducting order parameter \cite{salkola1997spectral, flatte1997localDefects, flatte1997localMagnetic, balatsky_impurity-induced_2006, meng2015superconducting, pershoguba2015currents, graham2017imaging, bjornson2017superconducting, theiler2019majorana}.
However, the coupling between the tip and the impurity is very weak and, therefore, the influence of the impurity on the tip superconductor is negligible, and thus we assume $\Delta_{\vect{i}, t} = \Delta_{0}$.
On the other hand, the effects of the impurity on the order parameter of the substrate $\Delta_{\vect{i},s}$ are significant and we calculate these effects self-consistently.

For the self-consistent calculation, we ignore the effects of the tip on the rest of the system by setting $t_t = 0$ and solve the gap equation for each lattice site in the substrate. 
For the $n$-th iteration of the self-consistent loop, the order parameter is calculated by \cite{gennes_superconductivity_2019, Black-SchaffePhysRevB.78.024504, bjornson_vortex_2013, reis_self-organized_2014, bjornson_probing_2015, awoga_disorder_2017, bjornson2017superconducting, theiler2019majorana, theiler_temperature_2025}:
\begin{equation}
    \label{eqn:gapequation}
\begin{split}
    \Delta_{\vect{i},s}^{(n)} &= - g/2 \left( \left \langle c_{\vect{i},\downarrow,s} c_{\vect{i},\uparrow,s} \right \rangle - \left \langle c_{\vect{i},\uparrow,s} c_{\vect{i},\downarrow,s} \right \rangle \right) \\
    &= - g/2 \sum_{\nu} f_T( E_\nu) \left( v^*_{\nu,\vect{i},\downarrow,s} u_{\nu,\vect{i},\uparrow,s} - v^*_{\nu,\vect{i},\uparrow,s} u_{\nu,\vect{i},\downarrow,s} \right),
\end{split}
\end{equation}
where $g$ is the superconducting interaction that we set to $g = 3.60\,t$ giving a bulk order parameter of $\Delta_{0} = 0.24\,t$. The Fermi-Dirac function $f_T$ is evaluated at the temperature $T$.
The expectation values $\left \langle c_{\vect{i}, \uparrow,s} c_{\vect{i}, \downarrow,s} \right \rangle$ are calculated from the corresponding particle-hole components of the eigenstates $u_{\nu,\vect{i},\sigma,s}$ and $v_{\nu,\vect{i},\sigma,s}$ for each eigenenergy $E_\nu$ calculated for the Hamiltonian including the local order parameter of the previous self-consistency step $\Delta_{\vect{i},s}^{(n-1)}$.
For the first iteration step we use the homogeneous order parameter distribution of $\Delta_{\vect{i},s}^{(0)} = \Delta_{0}$ or a previously calculated distribution with a similar impurity substrate coupling $t_s$.
We also test other starting points for $\Delta_{\vect{i},s}^{(0)}$, such as random distributions or homogeneous distributions with a lower order parameter than $\Delta_{0}$ but find reliable convergence to the same final $\Delta$.
As the convergence criterion for this self-consistent calculation, we stop the loop when the maximum change of any local value of $\Delta_\vect{i}$ between consecutive iterations is below $5 \cdot 10^{-5} \Delta_{0}$.

To solve the gap equation Eq.~\refeqn{eqn:gapequation}, for each iteration of the self-consistency loop, we use exact diagonalization.
Although in principle other methods could be used, such as a kernel polynomial method \cite{weise_kernel_2006, efficient2010covaci, efficient2012Nagai, bjornson_majorana_2016, mashkoori_impurity_2017, theiler2019majorana, mashkoori_identification_2020, lothman_efficient_2021}, we find it too inaccurate with the low number of polynomials needed to enhance performance over exact diagonalization.

%
\subsection{Josephson current through the impurity}
\label{subsec:josephson_current_through_imp_methods}
%


An applied bias voltage between the tip and the impurity will drop mainly between the tip and the impurity, due to the low coupling $t_t$.
This allows us to perform a gauge transformation that removes the phase difference between the tip and substrate order parameters and instead moves it into the description of the coupling Hamiltonian $V_{i, t}$ \cite{cuevas_hamiltonian_1996}:
\begin{equation}
\label{eqn:Hamil_coupling}
\begin{split}
    V_{i,t} &= - \sum_{\sigma} t_{t} \exp{\i \varphi(\tau)/2}  d^{\dagger}_{\sigma} c_{\vect{x}_0, \sigma, t} + \text{h.c.},
\end{split}
\end{equation}
where the time $\tau$ dependent phase difference $\varphi(\tau)$ includes the voltage $V$ between the tip and the substrate, given by:
\begin{equation}
\begin{split}
    \varphi\left(\tau\right) = \varphi_0 + \frac{2 \pi}{\phi_0} V \tau,
\end{split}
\end{equation}
with $\varphi_0$ being the constant phase difference between the tip and substrate order parameters and $\phi_0$ denotes the magnetic flux quantum.

To calculate the current between the tip and the impurity-substrate system, we start with the general expression for the tunneling current from the tip to the impurity \cite{cuevas_hamiltonian_1996},
\begin{equation}
    \label{eqn:general_current_expression}
\begin{split}
    I(\tau) = \frac{ie}{\hbar} \sum_{\sigma} &\left[ t_{t} \exp{\i \varphi(\tau)/2} \langle d_{\sigma}^\dagger (\tau) c_{\vect{x}_0, \sigma, t} (\tau) \rangle \right. \\
    &\left. - t_{t} \exp{-\i \varphi(\tau)/2} \langle c_{\vect{x}_0, \sigma, t}^\dagger (\tau) d_{\sigma} (\tau) \rangle \right].
\end{split}
\end{equation}
This expression can be evaluated using Keldysh Green's functions \cite{keldysh1964diagram}.
Generally, the current separates into two contributions \cite{cuevas_hamiltonian_1996, villas2020interplay}:
\begin{equation}
    I(\tau, \varphi_0, V) = I_{qp}(V) + I_J(\tau, \varphi_0, V).
\end{equation}
The quasi-particle contribution $I_{qp}$ describes single particle tunneling, which only comes into effect if the bias voltage is large enough to overcome the superconducting gap in the spectrum, i.e. $|V| > 2\Delta_{0}$. 
The Josephson current contribution $I_J$ describes the tunneling of Cooper pairs \cite{josephson1962possible}. We will focus on the dc-Josephson current at zero bias voltage in this work.

Using the fact that the coupling that drives the tunneling current is weak, the tunneling current can be expressed as \cite{cuevas_hamiltonian_1996, graham2017imaging}:
\begin{equation}
 \label{eqn:dc_josephson_current}
 \begin{split}
    I_J &= I_C \sin (\varphi_0), \\
     I_C &= \frac{4 e t_t^2}{\pi } \int \text{d}\omega f_T(\omega) \text{Im}[F_t^r(\omega) F_i^r(\omega)],
 \end{split}
\end{equation}
where $I_C$ denotes the critical Josephson current.
The electron charge is denoted by $e$ and $f_T$ is the Fermi function at $T$.
This expression only considers terms up to second order of $t_t$, but given that $t_t$ is assumed to be small, this is a good approximation.
The calculation relies on knowledge of the local retarded anomalous Green's functions $F_{\alpha}^r$ of both the tip at $\vect{x_0}$ and the impurity coupled to the superconducting substrate, but we drop the superscript $^r$ for ease of notation.
These are extracted from the corresponding elements of the Green's function matrix, which we define as $\hat{G}=\left[ (\omega + \i \delta) - \Hamil_{\alpha} \right]^{-1}$:
\begin{equation}
 \label{eqn:numerical_anomalous_greensfct}
    F_{\alpha}\left(\omega \right) = \sum_\sigma \frac{v^*_{\nu,\alpha \sigma} u_{\nu, \alpha\bar{\sigma}} }{\omega - E_\nu + \i \delta }.
\end{equation}
The parameter $\delta = 10^{-3}\,\Delta_{0}$ ensures numerical convergence and describes the experimental broadening of spectral features, commonly referred to as the Dynes parameter \cite{dynes_direct_1978}.
We choose its value to be in line with estimates from related experimental setups \cite{huang2020tunnelling}.
We find that apart from drastically increasing $\delta$, the exact value does not qualitatively change our results.
Furthermore, since the expected influence of the impurity on the tip is negligible, $F_t$ is assumed to be constant.
On the other hand, $F_i$ depends on $t_s$, as well as on the superconducting order parameter, which changes when self-consistency is achieved.
As a reference, we additionally calculate the current $I_0$, which denotes the Josephson current from the tip to a substrate unaffected by the impurity, while using the same coupling $t_t$.

\subsection{Josephson current to the substrate}
\label{subsec:methods_substrate_tunneling}

We also investigate the Josephson current flowing directly from the tip into the substrate, close to the impurity.
In JSTM experiments, this can be probed by scanning the tip across the substrate in the vicinity of an impurity \cite{randeria_scanning_2016, senkpiel2020single}.
This would make it possible to probe the extent of the YSR state \cite{graham2017imaging}.

The Josephson tunneling current from the tip to a lattice site $\vect{x}$ is calculated via Eq. ~\refeqn{eqn:dc_josephson_current}, where the anomalous Green's function of the impurity is replaced by the corresponding one of the site $\vect{x}$, i.e. $F_i \rightarrow F_\vect{x}$.
This effectively means changing the coupling term between the impurity and the tip to a coupling that connects site $\vect{x}$ in the substrate to the tip instead. 
The change in the Hamiltonian Eq. \refeqn{eqn:Hamil_coupling} is then:
\begin{equation}
    \label{eqn:Hamil_coupling_scanning}
    V_{t, i} \rightarrow V_{t, s}\left(\vect{x}\right) = - \sum_{\sigma} t_{\vect{x}}  c^{\dagger}_{ \vect{x}, \sigma, s} c_{\vect{x}_0, \sigma, t} + \text{h.c.},
\end{equation}
where $\vect{x}$ denotes the site in the substrate to which the STM tip has moved and $t_{\vect{x}}$ denotes the strength of the new coupling term between the tip and the substrate.
This new coupling $t_{\vect{x}}$ is tuned in experiments by the tip substrate distance.
However, it does not qualitatively change our results, as it only enters as a prefactor in Eq.~\refeqn{eqn:dc_josephson_current}, and thus we set it to $t_{\vect{x}} = t_t$ to make our results consistent with the tunneling current through the impurity.

\subsubsection{Additional channels}
\label{subsec:extra_channels_methods}

In JSTM experiments, the $\pi$-shift in the critical Josephson current $I_C$ is only measured indirectly \cite{karan_superconducting_2022}, by finding a shift in the switching current $I_S$.
In these experiments, tunneling events occur through several different channels between tip and substrate, with the measured tunnel current being the sum of these channels \cite{huang_quantum_2020, senkpiel2020single, karan_superconducting_2022}, where the main contribution is generally from the channel involving the magnetic moment.
The shift in $I_S$ has been attributed to the $\pi$-shift in this main channel.

These additional channels can originate from tunneling through other orbitals in the impurity \cite{scheer_signature_1998, cuevas_microscopic_1998} or tunneling directly to superconducting lattice sites.
The latter situation is the same as in Section~\ref{subsec:methods_substrate_tunneling} and we also expect a similar behavior if we add non-magnetic orbitals to the impurity Hamiltonian \refeqn{eqn:hamil_mag_imp}.
We further assume that the coupling terms for  these channels are weak and that the tunneling events in different channels are independent.
With these assumptions, we can use the same formalism as introduced in Section~\ref{subsec:methods_substrate_tunneling} and add the different current contributions from channels that couple to substrate sites closest to the impurity.
Note that the YSR state spreads from the impurity to these sites and thus also influences the Josephson current through the substrate channels.

\section{Results}
\label{sec:results}

Here we present the Josephson tunneling current from the tip to the magnetic impurity-substrate system and on how it is influenced by the change of the local superconducting order parameter, with special focus around the QPT. 
In Section \ref{subsec:energy_spectrum}, we start our discussion by showing the energy spectrum of the YSR states, the suppression of the local order parameter at $\vect{x}_0$, and the Josephson tunneling current directly through the impurity site. Then in Section \ref{subsec:results_scanning_stm} we present complementary results for the spatial dependence of the tunneling current to sites in the substrate.
This gives insights into how other transmission channels can behave and contribute the total tunneling current in experiments.

\subsection{Energy spectrum, local order parameter, and Josephson current through the impurity}
\label{subsec:energy_spectrum}

\begin{figure}%
\centering
\includegraphics[width=0.49\textwidth]{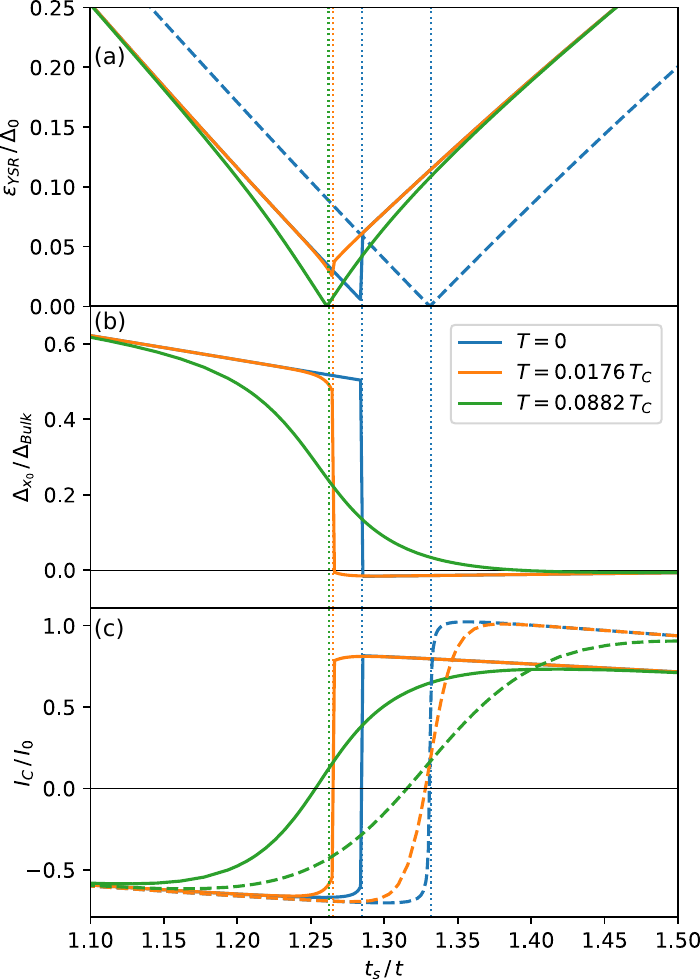}%
\caption{
(a) Energy of the unoccupied YSR state as a function of the coupling rate between the substrate and impurity $t_{s}$ for different temperatures $T$ relative to the critical temperature of the superconductor $T_C$.
(b) Local order parameter at $\vect{x}_0$ in the substrate.
(c) Critical Josephson current $I_C$.
Solid lines indicate self-consistent $\Delta_{\vect{i}}$, dashed line the constant $\Delta_{\vect{i}} =\Delta_0$ approximation.
Dotted vertical lines indicate the critical coupling $t_s = t_C$ at the QPT.
}%
\label{fig:spectrum_delta_Josephsoncurrent}%
\end{figure}


Before we start our discussion of the Josephson current through the impurity site, we discuss the energy spectrum of the YSR states, as it influences some of the behavior of the current.
Furthermore, the  crossing of the YSR states through zero energy in the spectrum is how the QPT can be detected in experiments \cite{hatter_magnetic_2015, farinacci_tuning_2018,  malavolti_tunable_2018, kezilebieke_observation_2019, huang_quantum_2020, karan_superconducting_2022, huang_universal_2023, karan_tracking_2024, uldemolins_hunds_2024} and is also how we identify the QPT in this work.
In Fig.~\ref{fig:spectrum_delta_Josephsoncurrent}(a) we show that the spectrum of the self-consistent order parameter solution depends on temperature, with the QPT occurring at somewhat lower values of $t_C$ (vertical dotted lines) for higher temperatures, in agreement with earlier results \cite{theiler_temperature_2025}.
At zero temperature (blue line), the YSR energy spectrum shows a gap and a jump at the QPT.
For low finite temperatures (orange), the gap widens.
This is due to the discontinuous change in the local order parameter for these temperatures, see Fig.~\ref{fig:spectrum_delta_Josephsoncurrent}(b)), where the local order parameter at the impurity site is heavily suppressed at the QPT to a small negative value, generating the $\pi$-shift in the local order parameter \cite{salkola1997spectral, flatte1997localDefects, flatte1997localMagnetic, balatsky_impurity-induced_2006, meng2015superconducting, pershoguba2015currents, graham2017imaging, bjornson2017superconducting, theiler2019majorana, theiler_temperature_2025}.
For higher temperatures (green) thermal excitations cause the discontinuous change in the local order parameter at the QPT to disappear \cite{theiler_temperature_2025}, which results in the YSR state energies smoothly crossing zero energy in the spectrum.
The YSR energy spectrum for a constant order parameter $\Delta_{\vect{i}} =\Delta_0$ (dashed line), on the other hand, is independent of temperature and always smoothly crosses through the Fermi level.


We show the critical Josephson current $I_C$ for the self-consistent solution (solid lines) and for the constant order parameter approximation (dashed lines) in Fig.~\ref{fig:spectrum_delta_Josephsoncurrent}(c).
For both solutions, we find that the critical current changes sign at the QPT, and for zero temperature, display a sharp change from positive to negative values in $I_C$.
This sharp discontinuity gets smeared out by temperature effects for the constant order parameter results, while the self-consistent solution retains the discontinuity for low, finite temperatures (orange line).
The latter is a direct result of the discontinuity of the YSR state energy at the QPT, as discontinuous changes in spectral features translate directly to the Josephson current through the impurity.
For temperatures where the discontinuity in the spectrum vanishes (green line), both solutions show a smooth transition of the current at the QPT.
Furthermore, we find an asymmetry between the critical current value in the weak and strong coupling regime close to the QPT, i.e.~the absolute value of $I_C$ is not symmetric around the QPT.
$|I_C|$ is generally higher for impurity substrate couplings $t_{s} > t_{C}$, albeit to a lower degree for the self-consistent order parameter solution.
However, this effect might still be hard to detect in experimental measurements \cite{karan_superconducting_2022}, as generally other transmission channels are also present that contribute to the Josephson current that would make such an offset hard to detect. 
A more distinct way in which the self-consistent solution differs from the constant order parameter approximation is that the $\pi$-shift in the critical Josephson current shifts with the QPT, with $t_{C}$ being lower at higher temperature.
However, this change in $t_C$ vanishes at higher temperatures, and thus we expect it to have only an effect at sufficiently low temperatures.

In summary, calculating the superconducting order parameter fully self-consistently, only changes the Josephson current very slightly compared to keeping the superconducting order parameter completely constant to its bulk value.
As such, the critical Josephson current cannot be used to measure the order parameter suppression and its $\pi$-shift at magnetic impurities.
The only notable influence of the order parameter suppression is that the current $\pi$-shift has a temperature dependence at the lowest temperatures, due to its influence on the YSR energy spectrum.

\subsection{Influence of local order parameter on Josephson tunneling in the vicinity of the impurity}
\label{subsec:results_scanning_stm}

\begin{figure}%
\centering
\includegraphics[width=0.49\textwidth]{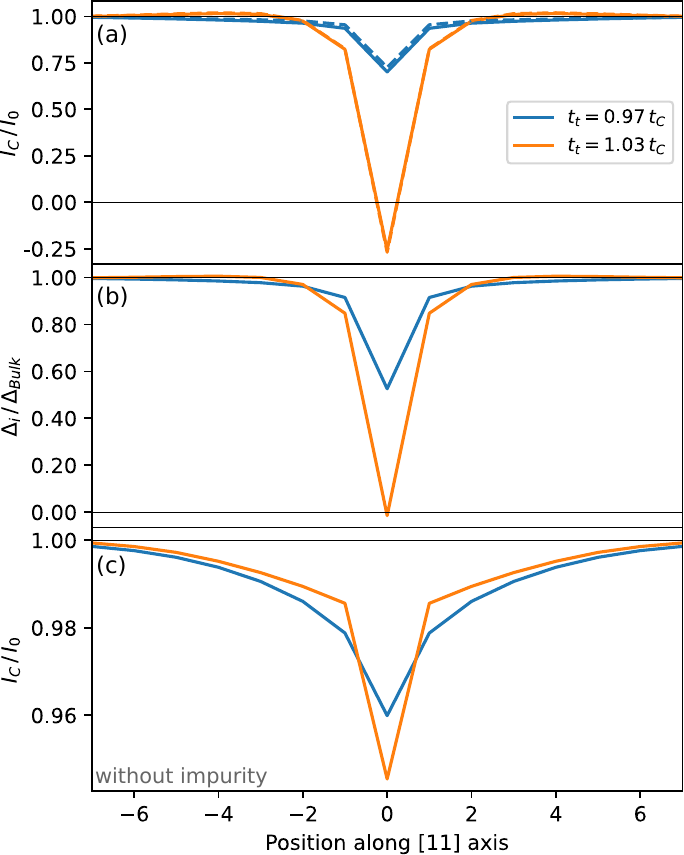}%
\caption{
Critical Josephson current $I_C$ for tunneling to the substrate (a) and local order parameter $\Delta_{\vect{i}}$ (b) along the $[11]$-axis intersecting $\vect{x}_0$.
(c) Critical Josephson current for tunneling into a substrate without the impurity present, but with the local order parameter distribution as in (b).
The blue (orange) lines represent two different coupling strengths between the substrate and the impurity, close to the QPT in the weak (strong) coupling regime.
Solid lines represent self-consistent order parameter solutions, dashed lines the constant order parameter approximation $\Delta_{\vect{i}} =\Delta_0$ at zero temperature $T=0$ in (a).
Note that the solid and dashed lines in (a) are very close and often overlap completely overlap.
}
\label{fig:scanning_across_substrate}%
\end{figure}

The YSR state is located mainly on the impurity site and therefore has the strongest effect on the Josephson tunneling current when tunneling through the impurity as discussed in the previous section.
However, the YSR state also spreads across the substrate, with a typical decay length depending on the Fermi velocity \cite{schlottmann1976spatial, salkola1997spectral,flatte1997localDefects, flatte1997localMagnetic, balatsky_impurity-induced_2006, graham2017imaging}.
This effect can potentially be observed in experimental settings that allow the tip to move spatially to substrate sites around the impurity \cite{randeria_scanning_2016, senkpiel2020single}.

In Fig.~\ref{fig:scanning_across_substrate}(a) we plot the critical Josephson current scanning through the impurity for couplings $t_s$ close to the critical coupling $t_C$ on both sides of the QPT.
As $t_C$ changes when self-consistency is enforced, we find that comparing the coupling values $t_s$ relative to the critical coupling $t_C$ gives the most consistent comparison.
We further note that the zero position in Fig.~\ref{fig:scanning_across_substrate} denotes the lattice site under the impurity, as that gives the most consistent comparison with neighboring sites. This site is experimentally accessible if the impurity is embedded in the substrate or through indirect channels tunneling.
We find the strongest variations in the tunneling current along the $[11]$-axis displayed in Fig.~\ref{fig:scanning_across_substrate}, but we note that we qualitatively find the same results in other directions.

Generally, Fig.~\ref{fig:scanning_across_substrate}(a) shows that for both the self-consistent order parameter solution (solid lines) and the constant order parameter approximation (dashed lines), the differences in the critical Josephson current are negligible.
Therefore, the Josephson current spatial profile is not influenced by the order parameter suppression.
For further comparison, we plot the self-consistent local order parameter Fig.~\ref{fig:scanning_across_substrate}(b) and find that its spatial profile is similar to the Josephson current.
However, this similarity cannot arise from the Josephson current merely following the order parameter profile, rather, it is a result of both quantities depending in a similar manner on the YSR state itself.
This relationship has been shown in Ref.~\onlinecite{graham2017imaging}, however, we in addition show that the order parameter suppression does not influence the Josephson current significantly.

To further solidify that the Josephson current spatial profile is not due to the local order parameter suppression, we remove the magnetic impurity, while artificially keeping the local order suppression in Fig.~\ref{fig:scanning_across_substrate}(b).
While this scenario is unlikely to occur in real world materials, as inhomogeneities are needed to disturb the local order parameter and these would inevitable create localized states changing the results, it still gives interesting insights.
Figure~\ref{fig:scanning_across_substrate}(c) shows the results.
There is now only a small suppression of about $5\%$ of $I_0$, which is a bit stronger in the strong coupling regime. 
Notably, the current profile does not match the local superconducting order parameter distribution, neither in its amplitude nor in its typical decay length.
Although the order parameter is heavily suppressed at the impurity site, even with a $\pi$-shift, and recovers within 3-5 lattice sites, the current is only slightly suppressed at the impurity site and decays slower to its bulk value. 
In summary, the Josephson current to the substrate and its spatial profile is not noticeably influenced by the local suppression of the superconducting order parameter. Instead, in agreement with the tunneling current through the impurity in Fig.~\ref{fig:spectrum_delta_Josephsoncurrent}, it must be the YSR state that directly influenced both the Josephson current and the superconducting order parameter.

\subsubsection{Additional tunneling channels}
\label{subsec:results_additonal_channels}

\begin{figure}%
\centering
\includegraphics[width=0.49\textwidth]{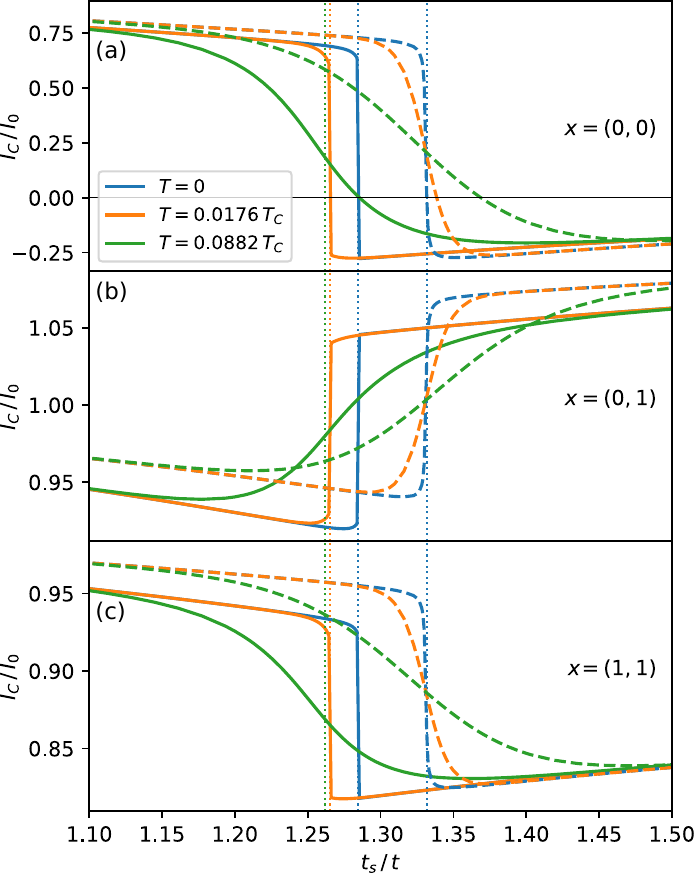}%
\caption{
Critical Josephson tunneling current directly to the substrate close to the impurity as a function of the impurity substrate coupling $t_s$ for the same temperatures as in Fig. \ref{fig:spectrum_delta_Josephsoncurrent}.
(a) Substrate site directly under the impurity $\vect{x_0}$, (b) nearest neighbor site to $\vect{x_0}$, and (c) next nearest neighbor site.
Solid lines represent self-consistent order parameter solutions, dashed lines the constant order parameter approximation.
Horizontal dotted lines indicate the critical coupling $t_C$ at the QPT for the respective temperatures.
}
\label{fig:current_additional_channels}%
\end{figure}
We finally show the Josephson tunneling current to sites close to the impurity as a function of the impurity substrate coupling $t_s$ across the QPT.
These results provide additional insight into other possible tunneling channels or pathways, which in an experimental setup contribute to the total measured current in experiments \cite{huang_quantum_2020, karan_superconducting_2022}.
In order to make the results comparable, here we use the same coupling $t_t$ for all channels, although this coupling would need to be fitted to experimental results for each channel.

In Fig.~\ref{fig:current_additional_channels}(a) we show the Josephson current to the site directly under the impurity ($\vect{x} = \vect{x}_0 =(0,0)$), (b) shows the nearest neighbor site ($\vect{x}=(0,1)$), and (c) the next nearest neighbor site ($\vect{x}=(1,1)$).
These results should further be compared to Fig.~\ref{fig:spectrum_delta_Josephsoncurrent}(c), which is the Josephson current through the impurity at $\vect{x} = \vect{x}_0 =(0,0)$.
All three channels show a shift in the critical Josephson tunneling current at the QPT, with some features similar to the tunneling current through the impurity site  in Fig.~\ref{fig:spectrum_delta_Josephsoncurrent}(c).
Interestingly, the direction of the critical Josephson current change at the QPT is not equal between the different sites.
The YSR state has a relatively strong spectral weight at position $x_0$ (Fig.~\ref{fig:current_additional_channels}(a)) and therefore the $\pi$-shift in the current is still observed at that site, however, with the opposite sign change compared to tunneling through the impurity in Fig.~\ref{fig:spectrum_delta_Josephsoncurrent}(c).
The other two sites (b,c) show no sign change, and the change in current at the QPT is about a magnitude smaller. We attribute these differences to the extension of the YSR states in the substrate. At the same time, we note that adding these channels to produce the full tunneling current, with weighted prefactors ($t_t$) depending on the experimental setup, will produce a way to indirectly measure the current $\pi$-shift and also a non-zero $I_C$  at the QPT, both measured experimentally \cite{karan_superconducting_2022}, with the latter requiring adding channels with no $\pi$-shift as in (b,c).

\section{Concluding remarks}
\label{sec:conclusions}

We investigate the effects of the local superconducting order parameter suppression by a magnetic impurity on the critical Josephson tunneling current flowing from a superconducting tip through the impurity or directly into the substrate.
At the QPT of the YSR state, the critical Josephson current through the impurity exhibits a $\pi$-shift in the superconducting current-phase relationship, i.e.~it has the opposite sign at the same superconducting phase difference on different sides of the QPT \cite{vecino_josephson_2003, martin2011josephson, graham2017imaging, kadlecova2019PracticalGuide}.
At the same time, local superconducting order parameter at the impurity site $\vect{x}_0$ also shows a $\pi$-shift or sign change at the QPT \cite{salkola1997spectral, flatte1997localDefects, flatte1997localMagnetic, balatsky_impurity-induced_2006, meng2015superconducting, pershoguba2015currents, graham2017imaging, bjornson2017superconducting, theiler2019majorana, theiler_temperature_2025}.
We find that both of these effects are induced by the presence of the YSR state.
However, the effects do not cause each other: the $\pi$-shift in the local order parameter does not directly cause the $\pi$-shift in the Josephson current. As a consequence, the Josephson current does not become a measure of the local superconducting order parameter near magnetic impurities.
In fact, we find that the local order parameter suppression has only very minor effects on the critical Josephson tunneling current.
The most notable effect is that the critical impurity substrate coupling $t_C$, at which the QPT occurs, is reduced when the suppression of the order parameter is accurately included and is further reduced with increasing temperature \cite{theiler_temperature_2025}.
The $\pi$-shift in the Josephson current follows this trend.

We further investigate the spatial dependence of the Josephson current and find that the current and the local order parameter display similar spatial profiles, in line with Ref.~\onlinecite{graham2017imaging}.
However, the direct effect of the local order parameter suppression on the Josephson current is still negligible.
The current (and order parameter) are instead mainly influenced by the extent of the YSR state.
In JSTM experiments, multiple tunneling channels are generally measured simultaneously, as Cooper pairs may take different paths between the tip and the substrate.
The resulting measured Josephson current then becomes a sum of these different tunneling contributions, with the channel through the impurity site being the main contribution \cite{karan_superconducting_2022}.
As a consequence, the spatial spread of the YSR state influences the total Josephson current, potentially even introducing a non-trivial behavior when different channels are combined. 

We finally note that in experiments the coupling parameter $t_s$, which determines if the system is in the weak or strong coupling regime with respect to the QPT, is not directly accessible.
Instead, the conductance of the tunneling junction is measured.
To couple our results with experimental observations, the conductance may be calculated in our model \cite{tersoff_theory_1985, chen_theory_1988}, by calculating the corresponding LDOS and fitting $t_s$ so that the QPT of the calculation results and the measurements align.
In addition, the values of $t_t$ and $t_{\vect{x}}$, for modeling the additional channels, would need to be fitted to experimental measurements with an exponential decay in the tip-sample distance.

\begin{acknowledgments}
The authors thank A.~V.~Balatsky and J.~C.~Cuevas for valuable discussions.
The computations were enabled by resources provided by the National Academic Infrastructure for Supercomputing in Sweden (NAISS), partially funded by the Swedish Research Council through grant agreement no.~2022-06725, and the Berzelius resource provided by the Knut and Alice Wallenberg Foundation at the National Supercomputer Centre (NSC).
The authors acknowledge the use of LLM-based tools to assist in language refinement during the preparation of this manuscript.
All data used to generate the figures in this work is publicly available under \cite{theiler_2026_18549215}.
\end{acknowledgments}

%

\end{document}